\documentclass{icrc2009}

\usepackage{graphicx}   
\usepackage[caption=false]{caption}    
\usepackage[font=footnotesize]{subfig} 
\usepackage{fixltx2e}
\usepackage{url}

\newcommand{\shorttitle}[1]%
{\markboth{Proceedings of the 31\MakeLowercase{$^{st}$} ICRC, {\L}\'{o}d\'{z} 2009}{#1} }
\newcommand{\etal}{\MakeLowercase{\textit{et al. }}} 

\newcommand{\degr}{$^{\circ}$}

\begin{document}
\title{A Topological Trigger System for Imaging Atmospheric-Cherenkov Telescopes}

\author{\IEEEauthorblockN{M. Schroedter\IEEEauthorrefmark{1},
			  J. Anderson\IEEEauthorrefmark{2},
			  K. Byrum\IEEEauthorrefmark{2},
                          G. Drake\IEEEauthorrefmark{2},
			  C. Duke\IEEEauthorrefmark{3},
			  J. Holder\IEEEauthorrefmark{4},
			  A. Imran\IEEEauthorrefmark{1},\\
			  A. Madhavan\IEEEauthorrefmark{1},
                          F. Krennrich\IEEEauthorrefmark{1},
                          A. Kreps\IEEEauthorrefmark{2},
			  A. Smith\IEEEauthorrefmark{2}
				}
                            \\
\IEEEauthorblockA{\IEEEauthorrefmark{1}Dept. of Physics and Astronomy, Iowa State University, Ames, IA, 50011, USA}
\IEEEauthorblockA{\IEEEauthorrefmark{2}Argonne National Laboratory, Argonne, IL 60439,USA}
\IEEEauthorblockA{\IEEEauthorrefmark{3}Physics Department, Grinnell College, Grinnell, IA 50112, USA}
\IEEEauthorblockA{\IEEEauthorrefmark{4}Department of Physics and Astronomy, University of Delaware, Newark, DE 19716, USA}
}

\shorttitle{Schroedter \etal Topological Trigger}
\maketitle

\begin{abstract}
A fast trigger system is being designed as a potential upgrade to VERITAS, or as the basis for a future array of imaging atmospheric-Cherenkov telescopes such as AGIS. The scientific goal is a reduction of the energy threshold%
\footnote{The energy threshold is defined as the peak in the differential trigger rate of a power-law spectrum with index -2.5.}%
 by a factor of 2 over the current threshold of VERITAS of around 130~GeV. The trigger is being designed to suppress both accidentals from the night-sky background and cosmic rays. The trigger uses field-programmable gate arrays (FPGAs) so that it is adaptable to different observing modes and special physics triggers, e.g. pulsars. The trigger consists of three levels: The level 1 (L1.5) trigger operating on each telescope camera samples the discriminated pixels at a rate of 400~MHz and searches for nearest-neighbor coincidences. In L1.5, the received discriminated signals are delay-compensated with an accuracy of 0.078~ns, facilitating a short coincidence time-window between any nearest neighbor of 5~ns. The hit pixels are then sent to a second trigger level (L2) that parameterizes the image shape and transmits this information along with a GPS time stamp to the array-level trigger (L3) at a rate of 10~MHz via a fiber optic link. The FPGA-based event analysis on L3 searches for coincident time-stamps from multiple telescopes and carries out a comparison of the image parameters against a look-up table at a rate of 10~kHz. A test of the single-telescope trigger was carried out in spring 2009 on one VERITAS telescope.
\end{abstract}

\section{Introduction}
We are designing a fast topological trigger system~\cite{NSS2008,Heidelberg2008} with the goal of reducing the energy threshold for detection of gamma rays by a factor of two below the threshold of current imaging atmospheric-Cherenkov telescopes (IACTs) around 130~GeV. This trigger will analyze in real time the trigger-maps across all telescopes in the array and thereby improve the rejection of background over the current generation of trigger systems. Aspects that ultimately determine how low the energy threshold can be, include not only the speed and sophistication of the trigger, but also the memory depth in the front-end electronics and the speed of the data acquisition system. This topological trigger may be part of a VERITAS upgrade proposal~\cite{Otte2009} or it may interface with a highly integrated camera on next generation IACTs, such as the Advanced Gamma-Ray Imaging System (AGIS).

\begin{figure}[th]
\begin{center}
\includegraphics[width=3.2in, angle=0.0,clip]{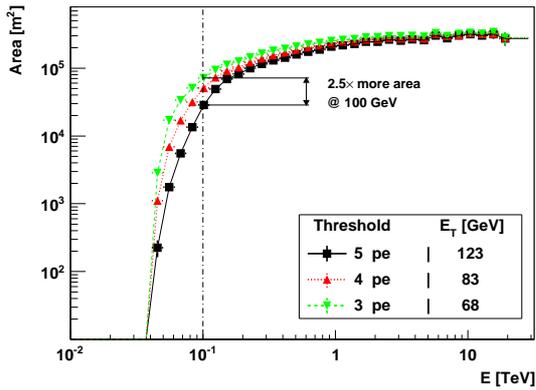}
\end{center}
\caption{Simulated gamma-ray collection area for VERITAS as a function of the trigger threshold of each photomultiplier tube (PMT). All gamma rays triggered at least 3 telescopes. By reducing the PMT threshold from 5~photoelectrons (nominal setting) to 3~photoelectrons, the collection area is increased significantly below 100~GeV and in turn lowers the energy threshold from 123~GeV to 68~GeV.  }
\label{fig:collection_area}
\end{figure}

The current generation of imaging atmospheric Cherenkov telescopes (IACTs) typically achieve gamma-ray collection areas of $10^5$~m$^2$ above 1~TeV. The gamma-ray collection area of VERITAS is shown in Fig.~\ref{fig:collection_area} at nominal trigger conditions (\emph{solid line}). The photomultiplier tube (PMT) thresholds are set at 5~photoelectrons (pe), resulting in an energy threshold at the trigger level, E$_t$, of E$_{t}\sim$112~GeV. A lower energy threshold could be achieved by decreasing the PMT thresholds. Fig.~\ref{fig:collection_area} shows two scenarios of lowered PMT thresholds that result in enhanced collection area below $\sim$100~GeV. However, these thresholds are not easily achievable because lowering the PMT threshold results in steeply rising data rates from background events: cosmic rays and night-sky background (NSB) light. 

\begin{figure}[th]
\begin{center}
\includegraphics[width=3in, angle=0.0,clip]{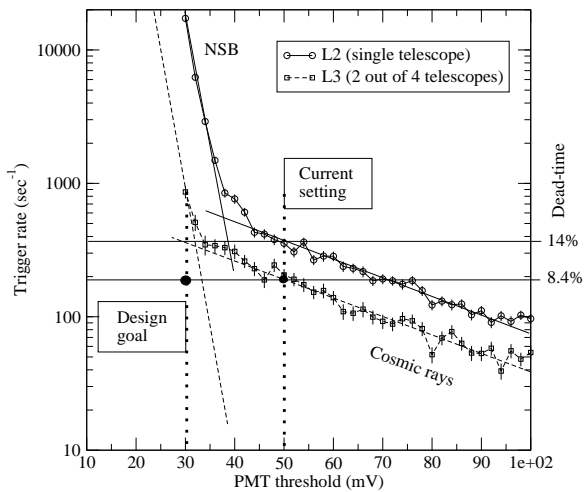}
\end{center}
\caption{ VERITAS trigger rates at the single-telescope level (\emph{solid line with circles}) and at the array level (\emph{dashed line with squares}). Only one of the four single-telescope curves is shown for clarity. The VERITAS array trigger requires at least two telescope to fire within 100~ns.  The fluctuations that are seen in the trigger rates are due due clouds moving through the field of view. To guide the eye, straight lines are drawn in the trigger regions dominated by cosmic rays and by night-sky fluctuations. The topological array trigger is being designed to allow lowering the threshold to $\sim$30~mV (about 3~pe) without increasing the dead-time.  }
\label{fig:array_bias}
\end{figure}

Rejection of background events with VERITAS is achieved through an efficient three-stage trigger~\cite{Holder2006}. At the primary level (L1), PMTs generate a trigger if at least $\sim$5~pe are detected within about 3~ns. Following the threshold requirement, a pattern-sensitive logic (L2) searches for 3 nearest-neighbor PMTs that trigger within a time span of about 7~ns. This camera trigger then sends a timing edge over optical fiber to a central station (L3). A trigger is issued by L3 only if at least 2 telescopes produced a trigger within 100~ns of each other. All parameters of the trigger setup are adjustable and were optimized for stable operation with less than 10\% dead-time under a variety of observing conditions, including moon light. The trigger rates of one of the telescopes and of L3 are shown in Fig.~\ref{fig:array_bias} as a function of the PMT threshold. It can be seen that cosmic rays are the main source of background at the current operating threshold of 50~mV (equivalent to 5~pe). However, if the PMT threshold is reduced to 3~pe, then a trigger would have to more efficiently reject both, NSB and cosmic rays. Fig.~\ref{fig:array_bias} indicates that at a threshold of 3~pe, the efficiency for rejection of NSB must be improved by $\sim10^4$ and that of cosmic rays by 70\%. Such a background reduction is possible if the stereo images (trigger maps) could be analyzed in real time at the array level. We are developing a trigger that implements a geometrical algorithm at the array level to reject more efficiently both, NSB and cosmic rays. In turn, the reduced background rate would allow the energy threshold to be lowered without increasing the telescope dead-time.

\section{Design Overview}
The topological trigger consists of high-speed camera-level triggers that transmit image parameters to the array trigger for real-time event classification. The design is implemented with field-programmable gate arrays (FPGAs) that allow the trigger to be adaptable to different observing modes (e.g. a ring-shaped acceptance region for low-energy pulsar studies) or concurrent special physics triggers (e.g. searching for short bursts of gamma rays). 

While the design is flexible in the specific deployment, the hardware specification were chosen so as to be compatible with VERITAS. In order to maintain a responsive system with a dead-time of less than 10

The first level of the topological trigger (L1.5) samples the discriminated camera signals at a rate of 400 MHz and searches for hits of three nearest neighbor pixels. There are three L1.5 cards per camera. The time window to detect coincidences is programmable to optimize the signal-to-noise ratio for detection of gamma rays. A fast L1.5 is essential since gamma-ray events as imaged by the telescope are of $\sim$5~ns duration. Currently, VERITAS is operating with a 7~ns coincidence window on each telescope. The increased speed alone makes it possible to lower the energy threshold without increasing the single-telescope trigger rate. To achieve the required level of timing accuracy across the camera, programmable delay lines are implemented on L1.5. This is necessary since nonuniform delays between pixels are introduced by cable lengths and variations in the high-voltage. The processing speed of L1.5 is well matched to the 2 ns samples recorded by the FADCs and facilitate an appropriate reduction of the energy threshold in the off-line analysis.

The hit pixels are then sent to a second trigger level (L2) that parameterizes the image shape and transmits this information along with the GPS time stamp to the array-level trigger (L3) at a rate of 10 MHz via a fiber optic link. The FPGA-based event analysis on L3 searches for coincident time-stamps from multiple telescopes and carries out a comparison of the image parameters against a look-up table at a rate of up to 10~kHz. We have identified an algorithm~\cite{Krennrich1995}, described below, that is relatively simple to implement on L3 and is suitable for gamma/hadron separation below 100~GeV.

\subsection{Parallactic Displacement}
\begin{figure}[th]
\begin{center}
\includegraphics[width=3in, angle=0.0, clip]{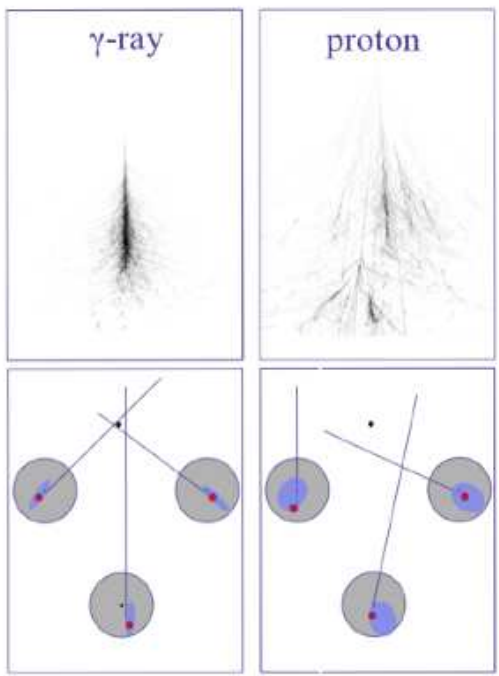}
\end{center}
\caption{Illustration of the images produced by gamma rays and cosmic rays in three telescopes. The shower core on the ground can be reconstructed by the intersection of the lines formed by the image centroid and the presumed gamma-ray source location. On average, proton showers exhibit a larger reconstruction error (parallactic displacement) than gamma rays. }
\label{fig:parallactic}
\end{figure}

The algorithm we intend to implement on L3 is based on the parallactic of shower-images seen from different points on the ground. As described in~\cite{Krennrich1995}, only a first moment-parameterization of each image is required and is practical to implement within the limited time of $\sim$10~$\mu$s available for data transmission and image processing. The method is illustrated in Fig.~\ref{fig:parallactic} for a gamma-ray and a cosmic-ray primary arriving on-axis. Gamma rays produce narrow and well-defined images in each telescope, while cosmic rays produce sub-showers and result in erratic distribution of the Cherenkov light between telescopes. The parallactic displacement results from the different view points of telescopes on the ground and displaces the image away from the camera center. By drawing lines through the image centroid and the camera center, pairs of telescopes produce an intersection. By projecting the intersecting lines on the ground, the impact location is reconstructed. The spread of the pair-wise intersecting points is smaller for gamma rays than for cosmic rays. The figure of merit, called \emph{parallaxwidth}, is defined by the coordinates of the intersecting points, $\vec{r}_i$, for each telescopes $i$, and the average over all such intersections, $\langle r_i \rangle$:  
\begin{displaymath}
\mathrm{parallaxwidth}=\sqrt{ \frac{ \sum_{i=1}^{n}(\vec{r_i}-\langle r_i \rangle)^2 }{n} }
\end{displaymath}

\begin{figure}[th]
\begin{center}
\includegraphics[width=3in, angle=0.0, clip]{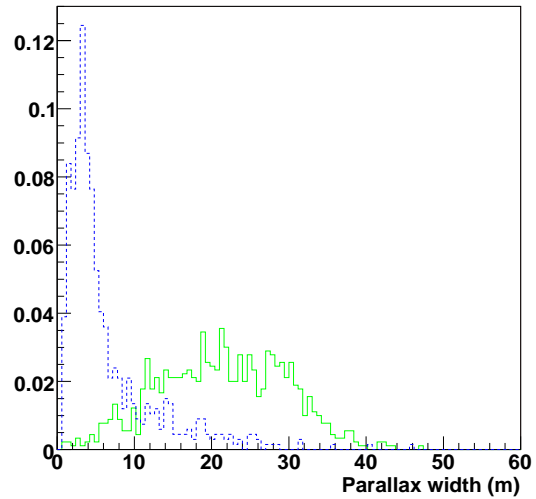}
\end{center}
\caption{ Comparison of the parallaxwidth distribution of 100~GeV gamma rays and cosmic rays, both arriving on-axis and from zenith. The parallaxwidth parameter is calculated for an array of 50 telescopes spaced 40~m apart. The area under the distributions are normalized with respect to each other. In the parameterization, only the locations of triggering PMTs with at least 3.5~pe are used. Cuts are applied on the minimum number of triggered PMTs ($>$5) and the pair-wise intersection angle ($>$30\degr).  }
\label{fig:parallax_dist}
\end{figure}

The efficiency of the \emph{parallaxwidth} parameter in separating gamma rays from cosmic rays is shown in Fig.~\ref{fig:parallax_dist} for an array of 50 telescopes. Gamma rays were simulated at 100~GeV, while cosmic rays were drawn from a power-law distribution with index -2.7 between 50~GeV and 10~TeV. The arrival direction for both, gamma and cosmic rays is on-axis, with telescopes pointed at zenith. Since cosmic rays are distributed isotropically, we are presenting a worst-case scenario for suppression of cosmic rays since off-axis cosmic rays will have even larger spread in the parallaxwidth parameter. The distribution on Fig.~\ref{fig:parallax_dist} results from a minimum of 3 telescopes participating with a PMT threshold of 3.5~pe. 

The NSB suppression expected with this topological trigger configuration is about $10^{-5}$ when compared to the VERITAS standard "2 out 4" array trigger.

\section{Status of Development}
A proof-of-principle trigger system is currently being developed by Iowa State University and Argonne National Lab. The hardware necessary for a complete single-telescope trigger has been produced and verified in the lab. The trigger resides in one VME crate containing three L1.5 cards, one L2 card, and thirteen Input/Output (I/O) cards. The I/O cards are necessary to test the topological trigger on a VERITAS telescope without disturbing night-time observations. For L3, we have purchased a PCI-based FPGA card from Faster Technology that features four fiber optical connections to the telescopes. The signal routing from the I/O cards to the L1.5 cards occurs over a custom backplane with gigabit connectors. The operation of the L1.5 cards has been verified at 400~MHz through direct pattern injection into the I/O cards. 

\begin{figure}[th]
\begin{center}
\includegraphics[width=3in, angle=0.0, clip]{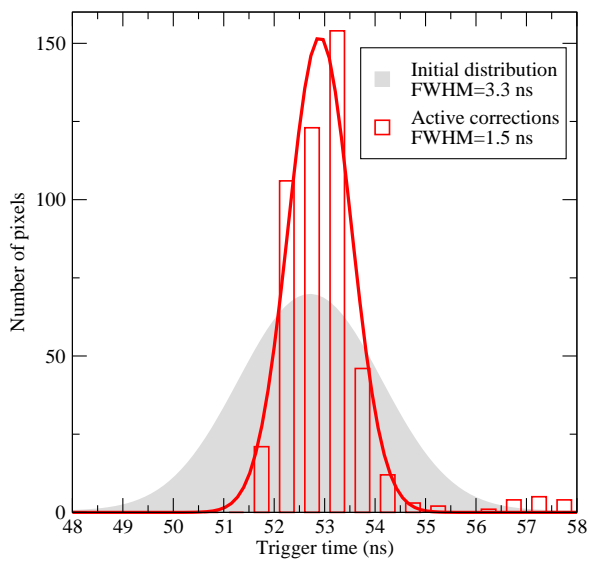}
\end{center}
\caption{Distribution of the relative trigger-hit arrival times inside the FPGA before (\emph{grey shaded area}) and after (\emph{solid lines and histogram}) the calibration with laser run. In principle the technique could reduce the FWHM to 0.16~ns once the VERITAS front-end delay modules are included in the calibration of the delays. }
\label{fig:delay}
\end{figure}

\begin{figure}[th]
\begin{center}
\includegraphics[width=3in, angle=0.0, clip]{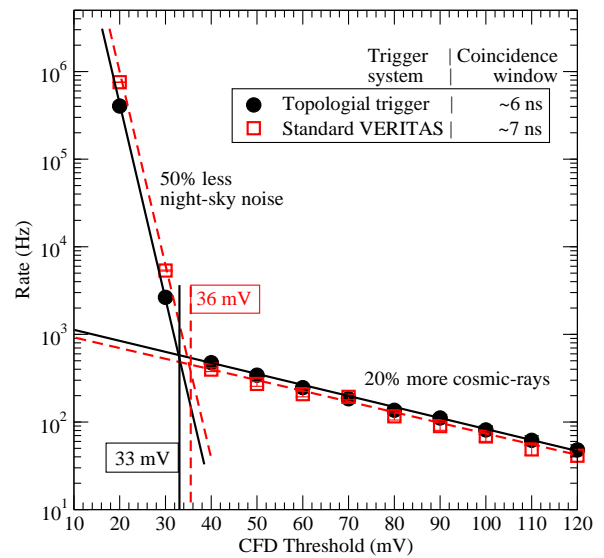}
\end{center}
\caption{Comparison of the trigger rate versus discriminator threshold for the topological trigger and the standard telescope trigger for VERITAS. Measurements were performed simultaneously under dark, moonless sky. The effective coincidence window of the topological trigger was 6~ns, while the standard trigger operates with a 7~ns wide window. The measured inflection point between the night-sky accidental hits and cosmic-rays is shifted from 36~mV to a slightly lower threshold of 33~mV for the topological trigger because of the shorter coincidence window. }
\label{fig:bias}
\end{figure}

Testing of the trigger was conducted in parallel with the standard trigger during late March 2009 on a single VERITAS telescope. The FPGA architecture includes adjustable front-end input delays to equalize the trigger arrival-times across all channels. Variations arise due to cable lengths and high-voltage settings. Delays up to 5~ns can be accommodated in the FPGA and are programmable with a precision of 0.078~ns. The measured timing distribution before and after compensating for the delay-variations is shown in Fig.~\ref{fig:delay}. Following the delay-line calibration, simultaneous measurements were taken of the trigger rate versus PMT threshold, shown in Fig.~\ref{fig:bias}. We observe a reduction in the night-sky accidentals by 50\% and an increase of the cosmic-ray rate by 20\%. The reduction in the night-sky background (NSB) rate is expected due to the faster coincidence logic (6~ns vs. 7 ns window). The increased cosmic-ray rate may be due to a larger trigger area at the edge of the field of view (10\%), and the fact that delay corrections are not applied to the VERITAS trigger inputs. The single-telescope trigger test was thus successful and we anticipate further reducing the NSB rate by shortening the coincidence window to 3~ns through firmware development.

\section{Acknowledgments}
We would like to thank the VERITAS collaboration for use of the telescope and support during the trigger test. This research is supported by Iowa State University, Argonne National Lab through the LDRD program, and the US Department of Energy through the Advanced Detector Research program (DEFG0207ER41497).

\end{document}